\documentclass[epjCONF,onecolumn]{svjour}
\usepackage{amsmath}
\usepackage{amsfonts}
\usepackage{amssymb}
\usepackage{graphicx}
\begin{document}
\title{Jet Quenching in Heavy-Ion Collisions}
\subtitle{The Transition Era from RHIC to LHC}
\author{Barbara Betz\inst{1}\fnmsep\thanks{\email{betz@th.physik.uni-frankfurt.de}}}
\institute{Institute for Theoretical Physics, Johann Wolfgang Goethe-University,
Frankfurt am Main, Germany}
\abstract{
A status report on the jet quenching physics in heavy-ion collisions is given as it appears 
after more than 10 years of collecting and analysing data at the Relativistic Heavy Ion 
Collider (RHIC) and $\sim$1.5 years of physics at the Large Hadron Collider (LHC). 
The (theoretical) predictions and expectations before the start of the LHC program are
contrasted with the most recent experimental results, focussing on the 
nuclear modification factor $R_{\rm AA}$, the elliptic flow $v_2$ of high-$p_T$ particles,
and on the problem of initial conditions.
} 

\maketitle
\section{Introduction}

In the year 2000, the physics program of the Relativistic Heavy Ion Collider 
(RHIC) started. For the first time, p+p, d+Au, and Au+Au collisions 
could be studied at identical centre-of-mass energies from 19.6 to 200 GeV using the
same detectors, BRAHMS, PHENIX, PHOBOS, and STAR.

The success of the RHIC program is mainly based on the fact that the results
obtained by the four experiments, summerized in a series of so-called white
papers \cite{whitebrahms,whitephenix,whitephobos,whitestar}, are in remarkable 
agreement with each other.

The main observations include: fast thermalization (indicated by a strong elliptic 
flow) \cite{Ollitrault:1992bk,Kolb:1999it}, low viscosity of the medium produced
(suggesting that it behaves like a ``nearly ideal fluid") 
\cite{Gyulassy:2004zy,Shuryak:2003xe,Romatschke:2007mq}, jet quenching (implying 
the creation of a dense and opaque system) \cite {Gyulassy:1990ye,Wang:1991xy},
strong suppression of the high-$p_T$ heavy-flavour mesons (the ``heavy-quark puzzle")
\cite{Adler:2005xv,Bielcik:2005wu}, and direct photon emission at high transverse 
momenta (confirming the scaling behaviour of hard processes) \cite{Adler:2005ig}.

However, after more than 10 years of RHIC physics, some fundamental questions still need to be clarified. 
What are the initial conditions of a heavy-ion collision? Is the medium created weakly 
or strongly-coupled? What is the process of fragmentation?

Two observables characterizing the medium are promising tools to help resolving those questions:
\begin{itemize}
\renewcommand{\labelitemi}{$\bullet$}
\item The nuclear modification factor, $R_{\rm AA}(p_T)$, para\-me\-trises the jet suppression and is 
defined as the ratio of jets produced in A+A collisions to the expectation for jets created 
in p+p collisions
\begin{equation}
 R_{\rm AA}(p_T) = \frac{dN_{\rm AA}/dp_T}{N_{\rm coll}dN_{\rm pp}/dp_T}\;.
\label{EqRAA}
\end{equation}
Here, $N_{\rm coll}$ is the number of binary collisions, a theoretical parameter 
that depends on the centrality of the collision and has to be 
calculated using a model describing the initial conditions. 
\item The elliptic flow $v_2$, defined as the second Fourier coefficient of the azimuthal particle emission,
\begin{eqnarray}
\frac{dN}{p_Tdp_Tdyd\phi} &=& \frac{1}{2\pi} \frac{dN}{p_Tdp_Tdy}
\left[1+2\sum\limits_{n=1}^\infty v_n (p_T,y,b)\cos(n\phi)\right]\,,
\label{Eqv2}
\end{eqnarray}
signals the creation of a medium whose expansion is determined by density gradients.
\end{itemize}

With the start of the Pb+Pb program at the Large Hadron Collider (LHC) in November 2010 
a new era began in the field of heavy-ion collisions. First results 
offered some surprises: While the center-of-mass energy per nucleon increases by a factor of ten 
as compared to RHIC energies and the particle multiplicity raises by a factor of
$\sim 2.2$ \cite{Aamodt:2010pb,ATLAS:2011ag,Chatrchyan:2011pb}, corresponding to a 30\% increase of 
temperature, the magnitude of the elliptic flow is similar to the results obtained at RHIC 
\cite{Aamodt:2010pa,ATLAS:2011ah,Collaboration:2012ta}.
Moreover, also the magnitude of jet quenching is surprisingly similar to RHIC for particles 
with $p_T>10$~GeV\footnote{Throughout this paper, natural units are applied with $c=\hbar=1$.} \cite{Otwinowski:2011gq,CMS:2012aa}.

In the light of these results, a status report on the jet quenching
physics in heavy-ion collisions shall be given. 
This report is not meant to review the underlying 
physics processes but to contrast the (theoretical) predictions and expectations 
before the start of the LHC program with the most recent experimental results.

The physics of jet quenching in heavy-ion collisions has many aspects and is explored
from several different angles. While both experiment and theory investigate
dijet imbalance \cite{Aad:2010bu,Chatrchyan:2011sx,Chatrchyan:2012nia,Qin:2010mn,Young:2011qx,Renk:2012cb}, 
$\gamma$-jet imbalance \cite{Morsch:2008zzb,Zhang:2009fg,ATLAS:Gamma-jet,CMS:Gamma-jet},
and fragmentation functions \cite{Morsch:2008zzb,Aad:2011sc,Yilmaz:2011zz}, 
the focus of this report is on the initial conditions, the nuclear modification factor $R_{\rm AA}(p_T)$, 
and the elliptic flow $v_2$ of high-$p_T$ particles since those observables already contain 
quite some information about the physical processes.

\begin{figure}[t]
\centering
  \includegraphics[scale = 0.45]{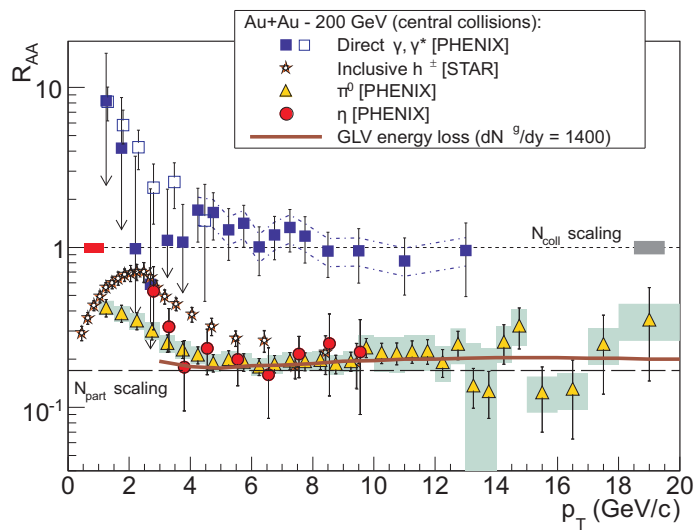}
  \caption{$R_{\rm AA}(p_T)$ measured in central Au+Au collisions at $\sqrt{s_{NN}}=200$~GeV, taken from 
  ref.\ \cite{ReviewDavid}, for direct photons \cite{Adler:2005ig}, $\pi^0$ \cite{Adler:2003pb}, $\eta$ 
  mesons \cite{Adler:2006hu}, and charged hadrons \cite{Adams:2003kv,Adler:2003au}, compared to 
  theoretical predictions of parton energy loss in a dense medium (solid brown
  curve) \cite{VitevGyulassy,VitevJet}. }
  \label{RAAPHENIX}
\end{figure}

\section{Parton Energy Loss and Initial Conditions}

The jet energy loss in heavy-ion collisions can either be described as multiple scatterings of
the parton \cite{bh,lpm,Baier:1996sk,Wiedemann:2000za,Qiu:1990xa,Arnold:2001ba}, 
specific for a weakly-coupled perturbative QCD (pQCD) medium, or using the string-theory inspired
Anti-deSitter/Conformal Field Theory (AdS/CFT) correspondence \cite{Maldacena:1997re}. For the latter,
the mathematical description of a parton stopped in a thermal medium is related to 
a string falling\footnote{To study a plasma at temperature $T$,
one needs to introduce a black hole in the AdS$_5$ geometry that has an event horizon at
$r_h$. By launching a D7 brane that spans from $r=0$ to $r=r_m$ (with $r_m > r_h$ for light quarks), 
the fundamental degrees of freedom (the quarks) are established. On this D7 brane, open strings represent $\bar{q}q$ pairs.
In the 5d geometry, these strings can fall towards the event horizon $r_h$. Thus, in case of light
quarks, open string endpoints will fall into the horizon. 
For a comprehensive summary on falling strings see {\it e.g.} ref.\ \cite{Ficnarnew}.}
into a $5$-dimensional black hole 
\cite{ches1,ches2,ches3,Arnold:2011qi,Ficnarnew}.

One of the interesting open questions is if {\it either} paradigm of a weakly or a
strongly-coupled medium can account for {\it both} RHIC and LHC observables. 

However, as shown in eq.\ (\ref{EqRAA}), the nuclear modification factor also depends on 
the initial conditions via the number of binary collisions. Two models are commonly used to describe those initial conditions. 
The Glauber model \cite{glauber} applies incoherent superpositions 
of p+p collisions while the ``Color Glass Condensate'' (CGC) 
\cite{cgc,Albacete:2010fs}, given {\it e.g.} by the Kharzeev-Levin-Nardi (KLN) model \cite{kln,kln2,kln3,adrian,dumitru}, 
takes saturation effects into account. Both models were shown to reproduce RHIC
results and exhibit large event-by-event fluctuations \cite{sorfluct,phogla,grassi,alver}.
However, they differ by their initial temperature gradients, 
their initial high-$p_T$ parton distribution, and the distance travelled 
by each parton. Thus, Glauber and CGC initial conditions should lead to 
a different opacity estimate and different magnitudes of the nuclear modification factor,
motivating a particular interest in the nuclear modification factor that
might convey information about {\it both} the initial conditions and the
jet-medium coupling.

\begin{figure}[t]
\centering
  \includegraphics[scale = 0.55,angle=270]{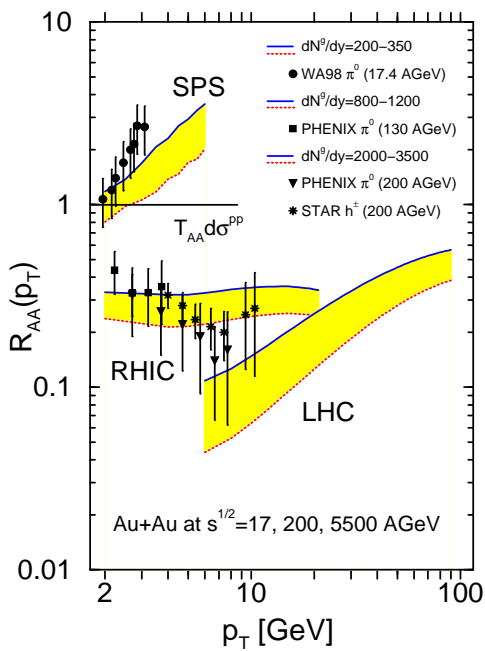}
  \caption{The nuclear modification factor $R_{\rm AA}$ as a function of $p_T$,
           taken from ref.\ \cite{Vitev:2002pf}, describing data from the 
           Super Proton Synchrotron (SPS), RHIC, and predicting the
	   trend for a pQCD-like energy loss at LHC conditions.}
  \label{VitevRAA}
\end{figure}

One of the first results of the heavy-ion program at the LHC \cite{Aamodt:2010pb}, the magnitude
of the particle multiplicity, indicated that the CGC initial conditions 
are disfavoured. However, a more recent work on CGC initial conditions \cite{Albacete:2010bs,Albacete:2011fw} 
shows that the centrality dependence of the hadron multiplicity both at RHIC and LHC is reproduced 
quite well if either $k_T$ factorisation is used or a formalism is applied that is based on the 
Balitsky-Kovchegov Equations \cite{Balitsky:1995ub,Kovchegov:1999yj} and includes the running-coupling 
corrections. Moreover, it was shown in ref.\ \cite{Schenke:2012wb} that quantum fluctuations
of color charges in a CGC-Glasma lead to higher moments of eccentricities 
[$e_n\sim v_n$ in eq.\ (\ref{Eqv2})] that are in remarkable agreement with results based on the Glauber model.  

Naturally, the differences in the initial conditions should not only be given in A+A
but also in p+A collisions. Since the proper description of the initial conditions
is still an open problem both at RHIC and LHC, the p+Pb runs at the LHC that are scheduled
for 2012 still have the potential of being a critical control experiment.

However, so far the predictions concerning the suppression in those p+Pb collisions,
that are parametrised by a suppression factor $R_{p{\rm Pb}}$, differ drastically. 
While it was shown in ref.\ \cite{Barnafoldi:2011px} that the
$R_{p{\rm Pb}}$ might allow for a clear disentangling of Glauber vs.\ CGC initial 
conditions, ref.\ \cite{Albacete:2010bs} identified a certain specification of
CGC initialisation for which a disentangling between different types of initial
conditions based on experimental data from the LHC will be impossible.

\section{RHIC Results and LHC Predictions}
\label{SecPredictions}

Figure \ref{RAAPHENIX} displays the nuclear modification factor of direct photons,
charged hadrons, pions, and $\eta$-mesons at RHIC energies. The photon 
$R_{\rm AA}(p_T>4\; {\rm GeV})\simeq 1$ is considered as proof that the direct 
photons measured do not interact with the medium created and can thus
be used as direct probes of the medium. On the other hand, the hadron suppression of $R_{\rm AA}(p_T)\sim 0.2$
indicates a rather strong jet-medium interaction that is similar for different
particle species (charged hadrons, pions, and $\eta$-mesons) and stays nearly flat up to
$p_T=20$~GeV.

One of the main questions in jet physics before the start of the LHC 
was if this $R_{\rm AA}(p_T)$ will stay flat or increase for larger $p_T$ and 
if this behaviour would allow to disentangle a weakly-coupled pQCD from
a strongly-coupled AdS medium.

Fig.\ \ref{VitevRAA} displays a synopsis of the $R_{\rm AA}(p_T)$ at different energies
\cite{lpm,Vitev:2002pf} applying the GLV formalism, ranging from the Super Proton 
Synchrotron (SPS) to the LHC, basically summarizing the results and predictions
for the nuclear modification factor of pions based on pQCD 
calculations \cite{Baier:1996sk,Wiedemann:2000za,Vitev:2002pf,Abreu:2007kv,Zakharov:1996fv,Baier:1996kr,Gyulassy:1999zd,Djordjevic:2003zk,Eskola:2004cr,Wicks:2007am,Loizides:2006cs,Qin:2007rn,Majumder:2007iu,Renk:2006pk,Bass:2008ch,Schenke:2009gb,Zapp:2009ud,Renk:2011gj,Chen:2011vt,Majumder:2011uk,Zapp:2011ek}.
While the Cronin enhancement \cite{Cronin:1974zm} dominates at SPS, shadowing and 
jet quenching cause a flat suppression pattern out to the largest $p_T$ at RHIC. 
For LHC energies however, the $R_{\rm AA}(p_T\sim 10\;{\rm GeV})$ is smaller than the one
at RHIC but rises for larger $p_T$. At those large transverse momenta ($p_T\gtrsim 10\;{\rm GeV}$)
the nuclear modification factor is completely dominated by the jet energy 
loss that depends on the particle multiplicity produced at LHC.

\begin{figure}[t]
\centering
  \includegraphics[scale = 0.45]{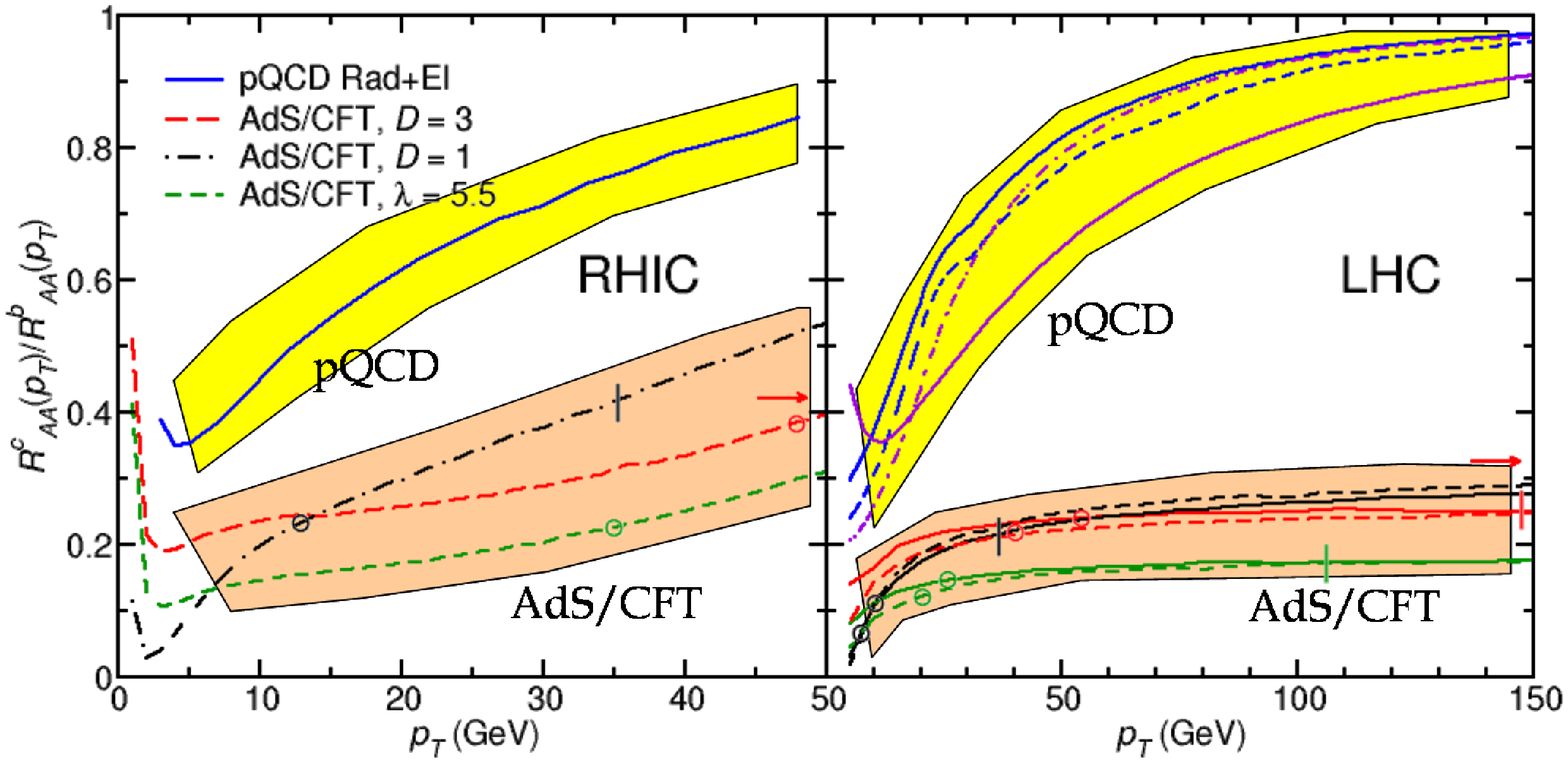}
  \caption{The ratio of a charm to a bottom nuclear modification factor as a function of $p_T$
    for RHIC (left panel) and LHC energies (right panel), comparing a pQCD vs.\ an AdS/CFT-inspired
    model fixing either the t'Hooft coupling $\lambda$ or the momentum diffusion coefficient 
    $D=2/\sqrt{\lambda}\pi T$ \cite{Horowitz:2007ui}. The (yellow and orange) bands display
    the cumulative uncertainties. The additional lines on the right-hand side correspond to different
    implementations of the initialization and the energy loss. For details see ref.\ \cite{Horowitz:2007ui}.}
  \label{WillRAA}
\end{figure}

While the nuclear modification factor at RHIC for light particles ({\it e.g.} pions) 
is in remarkable quantitative agreement with the measured data 
(see figs.\ \ref{RAAPHENIX}, \ref{VitevRAA}, and refs.\
\cite{Vitev:2002pf,Abreu:2007kv,Djordjevic:2003zk,Eskola:2004cr,Loizides:2006cs,Qin:2007rn,Majumder:2007iu,Renk:2006pk,Bass:2008ch,Schenke:2009gb,Renk:2011gj}),
the quenching of heavy quarks is significantly underpredicted
\cite{Dokshitzer:2001zm,Djordjevic:2005db,Wicks:2005gt}, an effect 
closely connected to the ``heavy-quark puzzle": In contrast to theoretical
predictions, the nuclear modification factor measured for heavy quarks suggests
that the magnitude of the energy loss for heavy quarks is similar to that one of light quarks.

However, such a heavy-quark energy loss can be determined using the AdS/CFT correspondence 
\cite{Armesto:2003jh,Moore:2004tg,Herzog:2006gh,Gubser:2006bz,Liu:2006ug,CasalderreySolana:2006rq,Gubser:2006nz,Horowitz:2007su,Horowitz:2007ui,Ficnar:2011yj}  
which also allows to explain the small viscosity to entropy ratio $\eta/s$ of the ``nearly perfect fluid"
\cite{Kovtun:2004de} at RHIC. This success motivated the application of string theory inspired models 
to the nuclear collision phenomenology, even though the AdS/CFT correspondence between string theory,
conformal Supersymmetric Yang-Mills (SYM) gauge theory, and non-conformal
QCD is still debated. Unfortunately, the AdS/CFT-based approaches for light-quark energy 
loss are much more complicated than for heavy quarks \cite{Ficnarnew,Chesler:2008uy},
limiting their application to heavy-ion phenomenology.

Using those AdS/CFT-based models to calculate the heavy-quark energy loss, the
$R_{\rm AA}(p_T)$ flattens at LHC energies as compared to RHIC energies
(see fig.\ \ref{WillRAA}). Thus, it was assumed that the slope of the 
nuclear modification factor as a function of $p_T$ could be an indicator for  
a pQCD weakly-coupled vs.\ an AdS/CFT-like strongly-coupled energy loss prescription. 

It should be stressed here that a recent work on falling string energy loss \cite{Ficnarnew}
has identified important corrections to the original works \cite{ches1,ches2} 
and indicates that also an AdS/CFT-like strongly-coupled energy loss prescription
could lead to an increasing $R_{\rm AA}(p_T)$.

One model that allows to test a weakly vs.\ a strongly-coupled
jet-medium interaction is the analytic geometric absorption model
introduced in refs.\ \cite{Horowitz:2011gd,Betz:2011tu,Betz:2012qq}
that can also be used to investigate Glauber vs. CGC initial conditions.
For this model, the energy loss per unit length, $dE/dx=dP/d\tau$ 
\begin{eqnarray}
\hspace*{-3ex}
\frac{dP}{d\tau}(\vec{x}_0,\phi,\tau)= 
-\kappa P^a(\tau){\tau^{z} T^{c=(z-a+2)}[\vec{x}_\perp(\tau),\tau]}\,
\label{GenericEloss}
\end{eqnarray}
is given as a function of proper time $\tau$ for a fixed jet 
rapidity $y$. The energy loss per unit length is characterized by the three exponents ($a,z,c$)
that determine the jet momentum dependence $P^a$, the path-length dependence 
$\tau^z$, and the local temperature power dependence $T^c(\vec{x}_\perp,\tau)$. 
The transverse jet path is $\vec{x}_\perp(\tau)=\vec{x}_0+\hat{n}(\phi)\tau$ from the 
production point $\vec{x}_0$ in direction of an azimuthal angle $\phi$ 
relative to the reaction plane. In case of radiative energy loss tomography,
the dimensionless effective medium coupling  $\kappa$ is proportional to $\alpha_s^3$
while it is proportional to $\kappa \propto \sqrt{\lambda_{tH}}\propto (\alpha_s N_c)^{1/2}$ 
in terms of the t'Hooft coupling at large $N_c$ in case of gravity-dual holography which allows
to quantify the magnitude of the jet-medium coupling at RHIC vs.\ LHC \cite{Betz:2012qq}.
$T(\vec{x},\tau)$ is the local temperature field of the QGP.

\begin{figure}[t]
\centering
  \includegraphics[scale = 0.55]{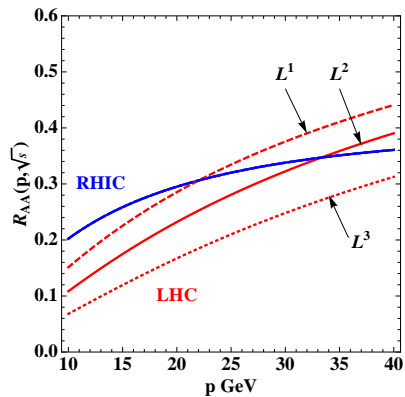}
  \caption{The nuclear modification factor as a function of $p_T$ 
   for RHIC and LHC energies, determined via the generic 
   absorption model in eq. (\ref{WillRAAEq}) for different estimates 
   of the path-length dependence ($L^1$, $L^2$, or $L^3$), assuming that the density 
   increases by a factor of 2.2 between RHIC and LHC \cite{Horowitz:2011gd}.}
  \label{WillABC}
\end{figure}

In the Bethe-Heitler limit $a=1$ and $z=0$, while in the deep Landau-Pomeranchuk-Migdal (LPM) 
\cite{Landau:1953um,Baier:1994bd} pQCD limit $a\sim0$ and $z\sim 1$. If $a=1$ and $z=2$, eq.\
(\ref{GenericEloss}) coincides with the model referred to as "AdS/CFT" in refs.\ \cite{Jia} while the 
heavy-quark string drag energy loss of conformal AdS holography \cite{ches1,ches2} 
is depicted by $a=1, z=0$. 

The scenario with $a=1/3$ and $z=1$ describes approximately both the 
pQCD and the AdS/CFT falling string cases \cite{ches1,ches2}. An $(E/T)^{1/3}$-energy 
dependence is numerically similar to the $\log(E/T)$ dependence predicted 
by fixed coupling pQCD energy loss in the range $10<E/T<600$ relevant both at RHIC and LHC energies. 
This power law is also predicted to be the lower bound of the power $a$ in the 
falling-string scenario in an AdS/CFT conformal holography \cite{Ficnarnew}.

While it has widely been assumed that varying $z=1$ to $z=2$ allows to interpolate between 
the weakly and strongly-coupled dynamical limits 
\cite{ches1,ches2,ches3,Arnold:2011qi,Betz:2011tu,Betz:2012qq,Jia,horojia,Adare:2010sp}, 
the recent work of ref.\ \cite{Ficnarnew} indicates that the necessary corrections to
refs.\ \cite {ches1,ches2} reduce the path-length power-law dependence from $z=2$ 
back to $z\approx 1$. Parametrically, this would make
pQCD and AdS/CFT descriptions virtually indistinguishable for light jets.

Applying ideal boost-invariant Bjorken hydrodynamics \cite{bjorken} 
and a uniform static plasma brick of thickness $L$,
the nuclear modification factor for a final momentum $p_f$ is given by \cite{Horowitz:2011gd}
\begin{eqnarray}
R_{\rm AA}(p_f) \approx \left[
1+\kappa\frac{(dN/dy)^{(2-a+z)/3}}{(L\;p_f)^{1-a}}
\right]^{\frac{a-n(p_f)}{1-a}}
\label{WillRAAEq}
\end{eqnarray}
which allows to directly access the spectral index $n(p_f)$.
The only parameter $\kappa$ needs to be determined by a fit to a reference
point usually chosen to be $R_{\rm AA}(p_T\sim 10\,{\rm GeV})$ \cite{Horowitz:2011gd,Betz:2012qq}.

Considering an energy loss with $a=1/3$ and ($z\in [1,3]$), the generic 
absorption model of eq.\ (\ref{WillRAAEq}) leads to an increase in the slope of the $R_{\rm AA}(p_T)$ 
for LHC as compared to RHIC energies (see fig.\ \ref{WillABC}) \cite{Horowitz:2011gd}, in line with the
 early pQCD calculations \cite{Vitev:2002pf}, if a reduced spectral index 
$n(p_f)$ is assumed at the LHC.
Moreover, fig.\ \ref{WillABC} shows that an increasing opacity (i.e.\ a larger value 
for the path-length dependence $z$) as well as a larger density at the LHC cause 
a lowering of the nuclear modification factor that actually falls below the 
$R_{\rm AA}^{\rm RHIC}(10 \lesssim p_T \lesssim 20 \;{\rm GeV})$.

\section{First results from the LHC}

\begin{figure}[t]
\centering
\begin{minipage}{7cm}
\includegraphics[scale=0.35]{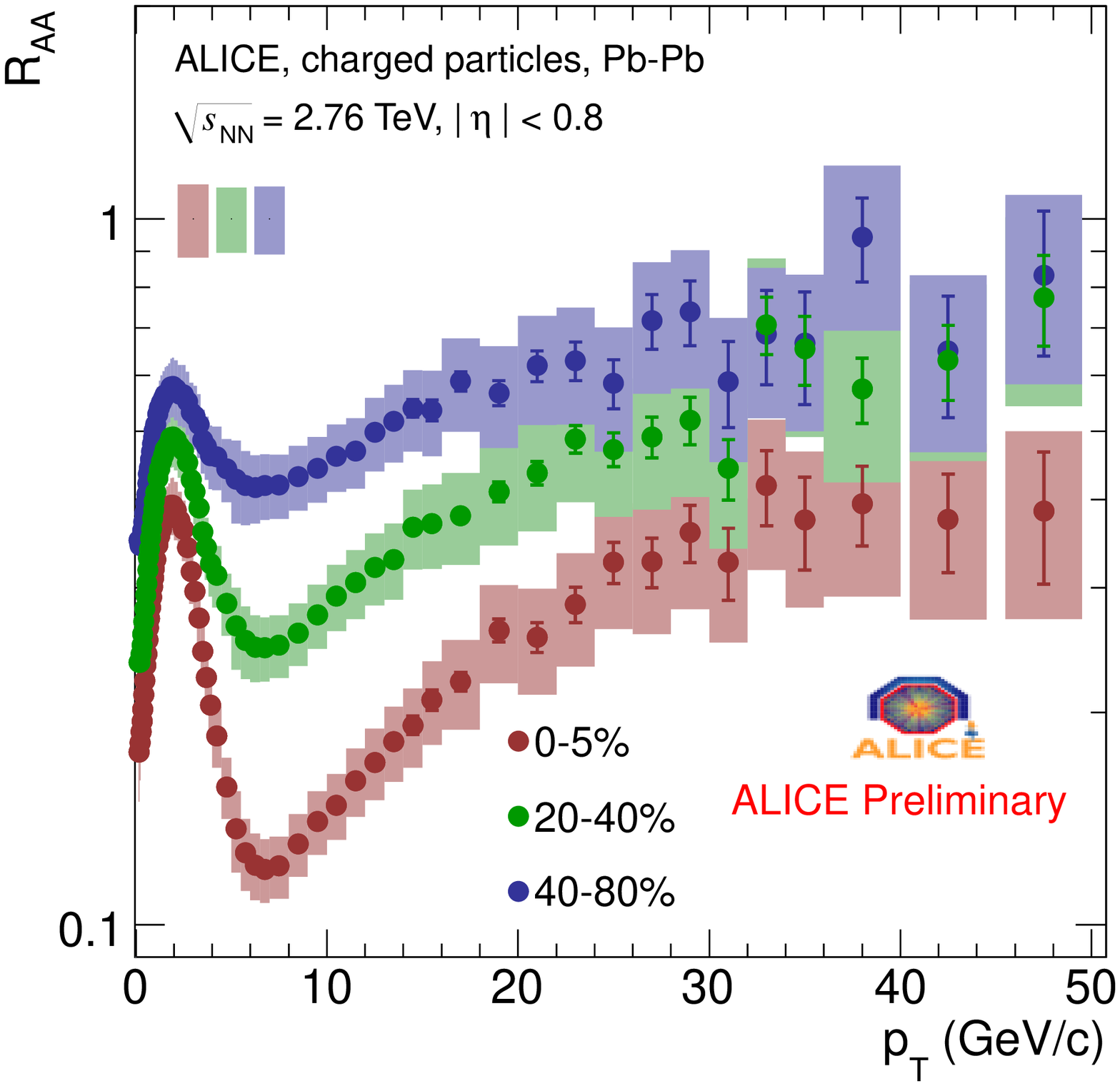}
\end{minipage}
\begin{minipage}{7cm}
\vspace*{0.5cm}
\includegraphics[scale=0.35]{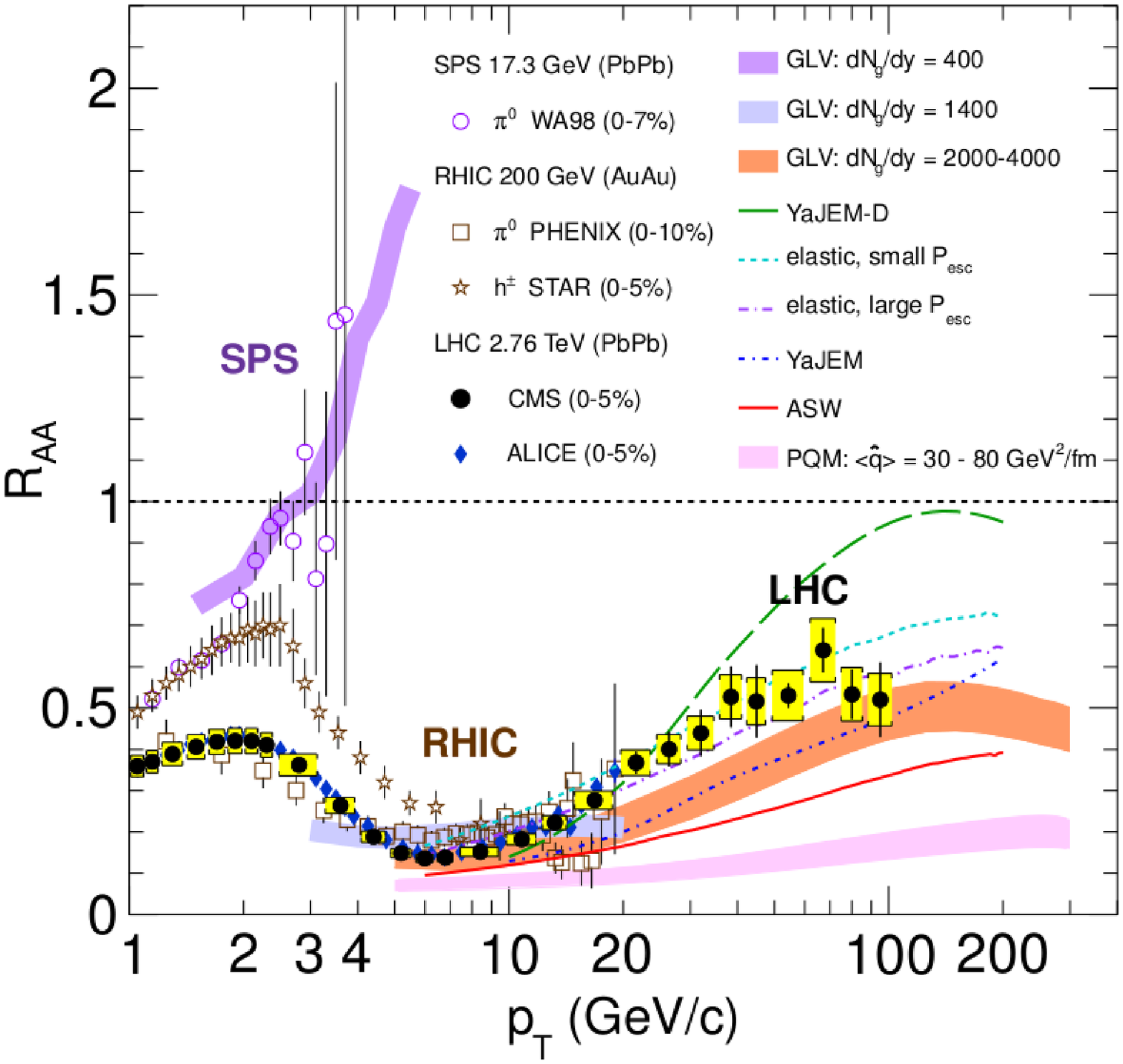}
\end{minipage}
  \caption{The nuclear modification factor at LHC energies as a function of $p_T$ for
   different centralities (left panel) \cite{Otwinowski:2011gq} and 
   compared to earlier results for lower energies at SPS and RHIC (right panel) \cite{CMS:2012aa}.}
  \label{LHC_RAA_results}
\end{figure}

\begin{figure}[b]
\centering
  \includegraphics[scale = 0.35]{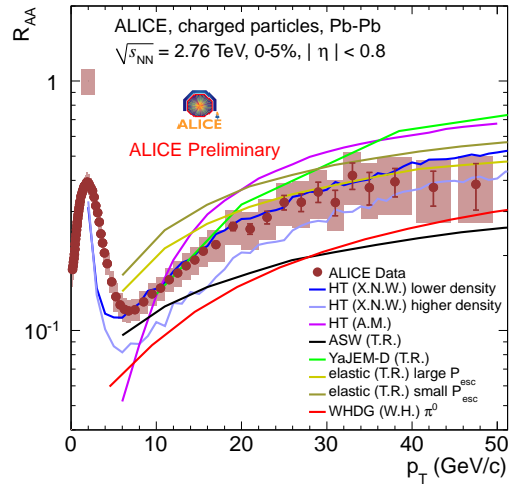}
  \caption{The nuclear modification factor as a function of $p_T$ at LHC energies as
   measured by the ALICE collaboration compared to different theoretical models \cite{Otwinowski:2011gq}.}
  \label{ALICE_compiled}
\end{figure}

First experimental results on the nuclear modification factor at LHC energies
showed (see fig.\ \ref{LHC_RAA_results}) that the $R_{\rm AA}$ increases with $p_T$, as originally proposed 
by pQCD \cite{Vitev:2002pf,Renk:2011gj,Chen:2011vt,Majumder:2011uk,Zapp:2009ud,Horowitz:2007ui} 
and recently also suggested for a revised AdS/CFT-inspired string model \cite{Ficnarnew}. Moreover, the 
$R_{\rm AA}(p_T)$ decreases with centrality and has a minimum at $p_T\sim 6-7$~GeV.

However, while the nuclear modification factor at LHC energies for $p_T<10$~GeV 
is significantly below the values at RHIC energies \cite{Otwinowski:2011gq,CMS:2012aa}
(cf.\ {\it e.g.} the right panel of fig.\ \ref{LHC_RAA_results}), RHIC and LHC results
for the $R_{\rm AA}(p_T)$ are in remarkable agreement for $p_T>10$~GeV \cite{CMS:2012aa}.

This ``surprising transparency" \cite{Horowitz:2011gd} contradicts   
LHC predictions based on density-de\-pen\-dent energy loss models 
\cite{Horowitz:2011gd,Betz:2011tu,Betz:2012qq,Zakharov:2011ws,Buzzatti:2011vt}, 
including GLV \cite{lpm}, ASW \cite{Armesto:2003jh}, PQM \cite{Dainese:2004te}, and YaJEM \cite{Renk:2011gj,Renk:2010zx}. 
Those models underpredict the measured data as seen 
in the right panel of figs.\ \ref{LHC_RAA_results} as well 
as in fig.\ \ref{ALICE_compiled}.

\begin{figure}[t]
\centering
\includegraphics[scale=0.45]{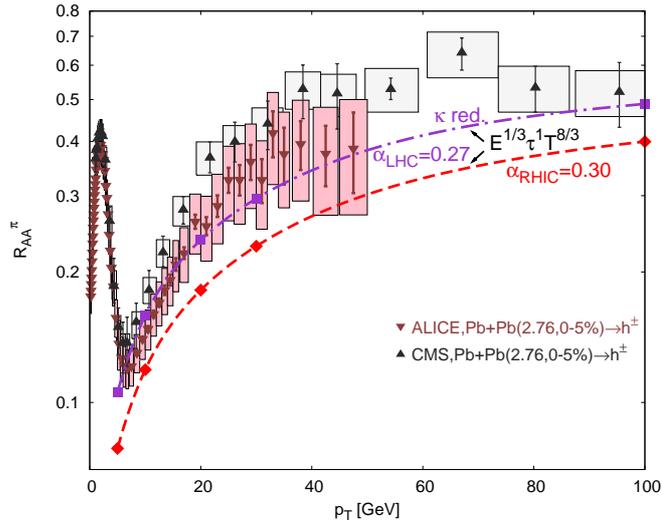}
  \caption{The nuclear modification factor as a function of $p_T$ at LHC energies for 
	different values of the coupling constant calculated using the generic energy-loss model    
        in ref.\ \cite{Betz:2012qq}.}
  \label{Us_RAA_results}
\end{figure}

To solve this puzzle, different approaches are chosen, considering {\it e.g.} different
densities for RHIC and LHC \cite{Chen:2011vt,Buzzatti:2011vt}, or adding
an escape probability of a parton without any medium interaction (YaJEM-D) 
\cite{Renk:2011gj}. Other recent studies \cite{Betz:2012qq,Pal:2012gf,Buzzatti:2012},
including one based on the generic energy loss introduced in eq.\ (\ref{GenericEloss})
for $a=1/3$ and $z=1$, showed that a plausible moderate reduction of the pQCD coupling 
[$\kappa\sim\alpha_s$ in eq.\ (\ref{GenericEloss})] due to slow running (creeping) of 
the coupling above the deconfinement temperature at LHC leads to a better description 
of the data as shown in fig.\ \ref{Us_RAA_results}.

However, considering a falling-string scenario 
the effective reduced jet-medium coupling $\kappa\propto \sqrt{\lambda}$ would imply a rather large reduction of
$\lambda_{\rm LHC}$ by a factor $\sim 2-4$ relative to RHIC \cite{Betz:2012qq}. It is not yet clear
if current non-conformal holographic models are consistent with such a 
strong variation (see for example, refs.\ \cite{Ficnar:2011yj,Mia:2011iv}), suggesting that
pQCD-based models might be favoured. Complementary to those results, the analysis for jet 
asymmetry and energy imbalance \cite{Aad:2010bu,Qin:2010mn}, as done {\it e.g.} using MARTINI 
\cite{Young:2011qx}, will provide further constraints on the jet-medium coupling.

The advantage of the above mentioned generic absorption model ($dE/dx=-\kappa E^a x^z T^c$) 
\cite{Betz:2012qq} introduced in section \ref{SecPredictions} is that it can 
not only analytically interpolate between weakly-coupled tomographic and strongly-coupled 
holographic jet energy loss models but also easily mimic different prescriptions for the 
energy dependence. 
An important result is that the class of geometric optics models with $a=1$, 
that had been very successful in describing the RHIC
data \cite{Jia,Adare:2010sp}, has to be excluded at the LHC since it does not reproduce 
the $p_T$-dependence of the nuclear modification factor at LHC \cite{Betz:2012qq}.

\section{Open Questions}

While LHC data seem to suggest that pQCD tomography is either favoured over AdS/CFT holography for $p_T>10$~GeV
or might even be indistinguishable from string inspired models for light quarks \cite{Ficnarnew,Betz:2012qq}, there are still 
open puzzles for both pQCD tomography and AdS/CFT holography, including the intermediate to high-$p_T$ 
elliptic flow \cite{Betz:2012qq,Horowitz:2011cv} and the heavy-quark jet quenching 
\cite{Wicks:2005gt,Adler:2005xv,Bielcik:2005wu,Noronha:2010zc,Djordjevic:2006kw}. 

Ref.\ \cite{Noronha:2010zc} shows that leading order AdS/CFT holography with a common
large t'Hooft coupling of $\lambda\sim 20-30$ may simultaneously describe the elliptic flow of bulk hadrons
as well as the nuclear modification factor of heavy-quark jet fragments. But it also predicts a much stronger 
suppression for charm particles. However, including dynamical multi-scattering in the
pQCD-based GLV model \cite{Buzzatti:2011vt}, this oversuppression is compensated. This might help solve the heavy-quark puzzle. 
It should be stressed that among the models computing both, the nuclear modification factor identified with D mesons {\it and} 
the $R_{\rm AA}$ of light quarks \cite{ALICE:2012ab}, only two models \cite{Buzzatti:2011vt,Sharma:2009hn}
describe both experimental observables reasonably well.

On the other hand, the $v_2$ puzzle still prevails at an intermediate $2\;{\rm GeV}<p_T<10\;{\rm GeV}$.
As shown in refs.\ \cite{Betz:2012qq,Horowitz:2011cv}, a geometric optics model with $a=1/3$ and $z=1$ [cf. 
eq.(\ref{GenericEloss})] describes the high-$p_T$ elliptic flow of particles with $p_T>10\;{\rm GeV}$
well but it cannot reproduce the $v_2$ data for $2\;{\rm GeV}<p_T<10\;{\rm GeV}$. This indicates that there is 
a clear ``intermediate band" for which a consistent description
of the nuclear modification factor and the elliptic flow at both RHIC and LHC energies
is still lacking.

All these findings together stress that any theory investigating high-$p_T$ data for both RHIC and LHC needs to be tested for 
the nuclear modification factor of light and heavy quarks as well as for the high-$p_T$ elliptic flow in
order investigate if this theory provides a coherent description of the underlying physics. 

Certainly, the current understanding of the physics at RHIC and LHC will be further tested
with the new runs scheduled at both RHIC and LHC as well as with the upcoming analyses of already collected
experimental data. A crucial test for all present theoretical models of jet energy loss will be
the release of data with $p_T>60\;{\rm GeV}$.

\section*{Acknowledgements}
The author acknowledges support from Alexander von Humboldt foundation and thanks
M.\ Gyulassy, G.\ Torrieri, J.\ Noronha, A.\ Buzzatti, A.\ Ficnar, W.\ Horowitz, J.\ Jia, J. Liao, and G.\ Roland
for fruitful discussions.

\end{document}